# PIP-II PROJECT PROGRESS AND CHALLENGES*

O. Napoly†
Fermi National Accelerator Laboratory, Batavia, IL, USA

*Abstract*

The Proton Improvement Plan-II (PIP-II) project is under construction to provide the Fermilab Accelerator Complex with a new H⁻ superconducting radiofrequency (SRF) linac and beam transfer line delivering 800 MeV protons at Booster injection at double energy and double intensity ($6.7\times10^{12}$ protons per 20 Hz pulse) in comparison with the present Linac. The design goals of PIP-II Linac are described, including delivering 1.2 kW beam power on Long Baseline Neutrino Facility (LBNF) target, continuous wave (CW) operation compatibility and future upgrades. The progress of the Linac construction. Finally, the challenges of operating the SRF Linac to drive the Fermilab Booster are addressed.

## PIP-II DESIGN GOALS

The Proton Improvement Plan-II (PIP-II) project [1] is an upgrade of the Fermilab accelerator complex whose main goals are to deliver 1.2 MW power on target from a beam of 120 GeV protons and to support the current 8 GeV program at Fermilab including the Mu2e and short baseline neutrinos experiments [2]. The LBNF beam line will extract protons from the Main Injector (MI) rapid cycling synchrotron (RCS) at 1.2 s cycle period which includes accumulation and storage of 12 8-GeV proton batches from Booster RCS cycled at 20 Hz in the Recycler Ring (RR) followed by transfer to MI and acceleration to 120 GeV. In parallel with the MI acceleration, another 12 Booster batches are distributed to various 8 GeV neutrino and muon experiments. Hence the total beam power delivered by the Booster will be in excess of 160 kW corresponding to $6.5\times10^{12}$ protons per 20 Hz pulse. Reducing the MI acceleration time will provide a path to LBNF power up to 2 MW with the identical Booster batch input and dropping the proton yield to the 8 GeV physics program.

The main elements of this upgrade (Fig. 1) are an 800 MeV superconducting radiofrequency (SRF) linac, named Linac2, accelerating $6.7\times10^{12}$ $H^-$ at 20 Hz repetition rate and a beam transfer line (BTL) delivering the $H^-$ beam to a new Booster injection region including an $H^-$ foil stripping system. The main parameters of the Linac2 are listed in Table 1. The maximum beam power at 800 MeV is 17.6 kW. This new facility will double the proton energy and intensity in comparison with the present 400 MeV $H^-$ Booster room-temperature injector, aiming at a yield of $3.3\times10^{21}$ protons/year assuming 76% uptime (44 weeks at 90% reliability).

Table 1: Linac2 Main Parameters

| Linac parameters | |
|---|---|
| Beam energy | 800 MeV |
| Beam pulse length | 550 µs |
| Repetition rate | 20 Hz |
| Pulse average current | 2 mA |
| Fundamental RF frequency | 162.5 MHz |
| Cryogenic power | 2.5 kW @ 2 K |
| Beam duty factor | 1.1 % |
| RF duty factor | 100 % |
| Linac/BTL length | 267/308 m |

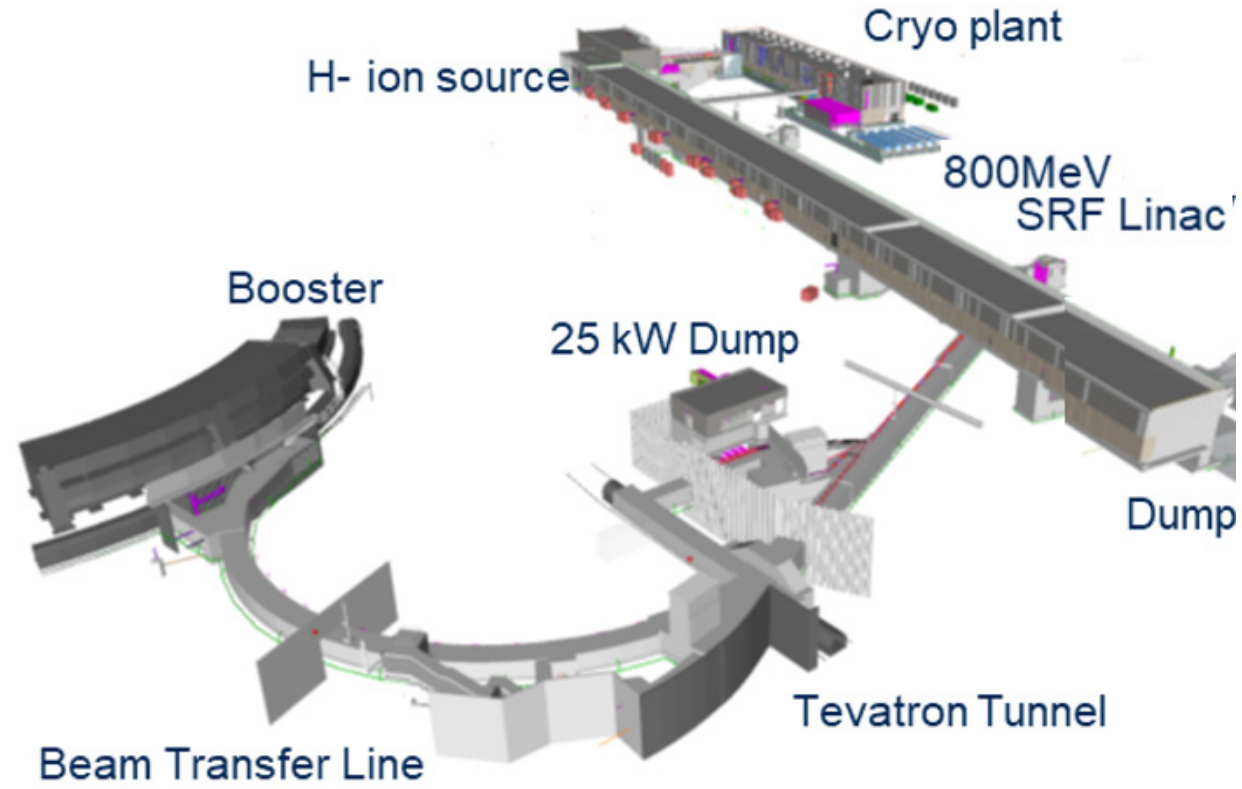

Figure 1: Schematic layout of PIP-II facility.

## THE PIP-II FACILITY

### *The Warm Font End*

The WFE starts with two 30 kV H- ion sources generating up to 15 mA current, followed by (Fig. 2):

- the low-energy beam transport (LEBT) section including a DC chopping system and a ±30° dipole which selects the beam from either of the two H- sources,
- a 4.43 m long radio-frequency quadrupole (RFQ) of 162.5 MHz RF frequency that brings the kinetic energy to 2.1 MeV, and
- a medium-energy beam transport (MEBT) section including 4 RF buncher cavities, a travelling-wave chopping system and a 21 kW-rated beam absorber.

The WFE is located in the injection high bay building (HBB) which includes all the WFE power supplies and utilities. However, the last WFE girder supporting two focusing triplets and the 4th RF buncher is in the linac tunnel with the upstream beam line traversing the HBB to tunnel concrete wall.

†napoly@fnal.gov

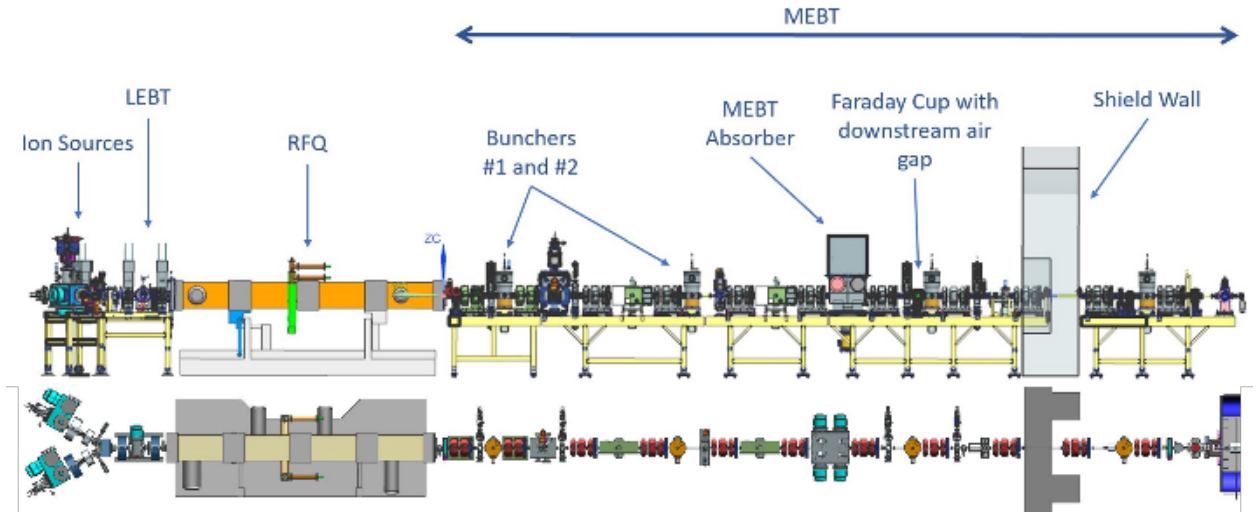

Figure 2: Layout of the PIP-II warm front end.

A gate valve and a differential pumping insert separate the WFE vacuum from the UHV and particle free vacuum section of the SRF linac. A fast-acting gate valve provides additional protection for the cryomodule string in case of a beamline vacuum leak. The same setup will be installed on the downstream end of the SRF linac.

## *The SRF Linac*

The linac includes five flavours of SRF cavities with increasing Lorentz β factor along with the increasing beam energy (Fig. 3). Cryomodules are assembled from these cavities in the following manner:

- The β=0.11 half-wave resonator at 162.5 MHz RF frequency populates one HWR cryomodule with 8 cavities and 8 interleaved superconducting solenoids (8×SC). The nominal cavity gradient is 9.7 MV/m.
- The β=0.22 single-spoke resonator (SSR) at 325 MHz RF frequency populates two SSR1 cryomodules with 8 cavities and 4 superconducting solenoids (4×CSC). The nominal cavity gradient is 10.0 MV/m.
- The β=0.48 SSR at 325 MHz RF frequency populates seven SSR2 cryomodules with 5 cavities and 3 superconducting solenoids (SCCSCCSC). The nominal cavity gradient is 11.5 MV/m.
- The β=0.61 elliptical resonator at 650 MHz RF frequency populates nine LB650 cryomodules with 4 cavities. The nominal cavity gradient is 16.9 MV/m.
- The β=0.92 elliptical resonator at 650 MHz RF frequency populates four HB650 cryomodules with 6 cavities. The nominal cavity gradient is 18.8 MV/m.

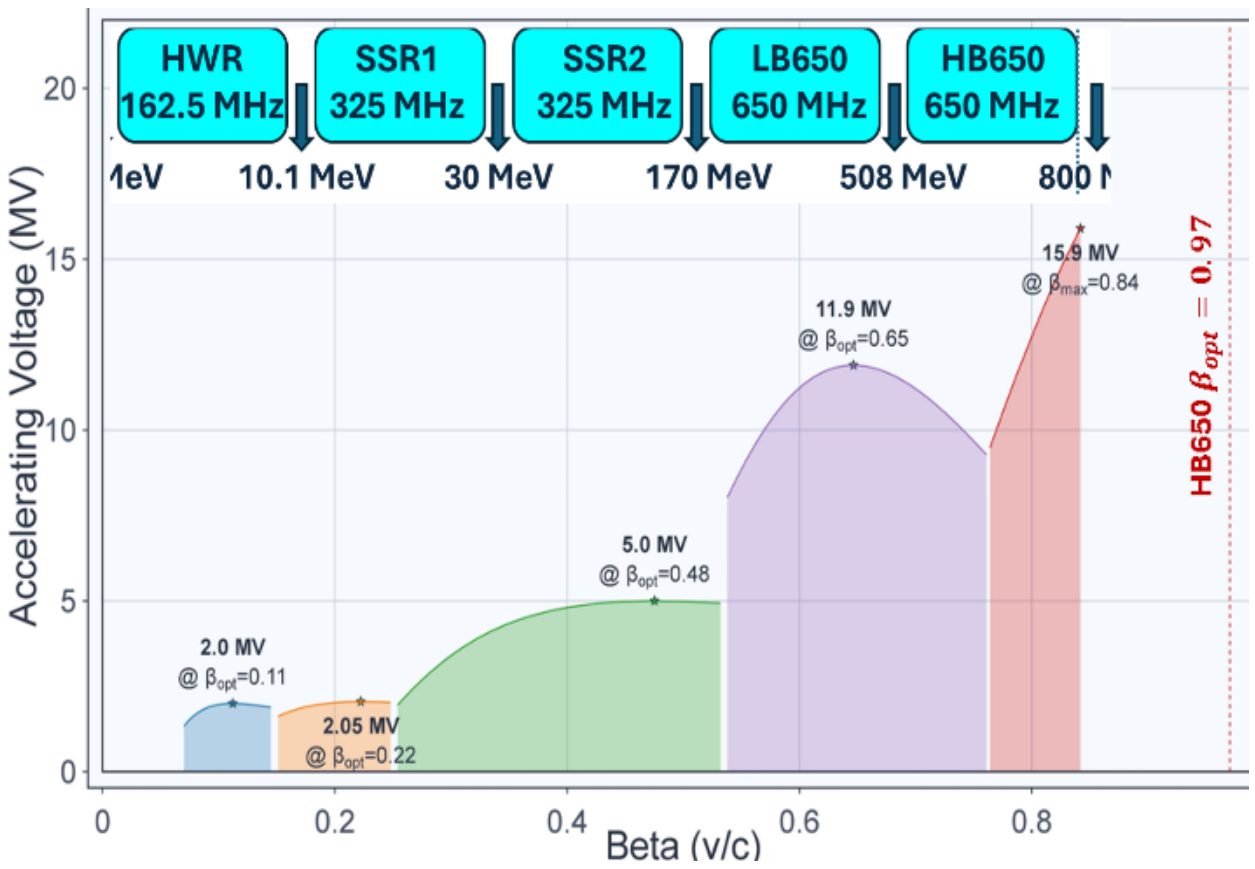

Figure 3: Nominal accelerating voltage of SRF cavities for increasing beam energy ranges of cryomodule types.

In the LB650 and HB650 sections beam focusing is provided by quadrupole doublets in warm insertions in front of every cryomodules.

In total Linac2 includes 23 cryomodules and 119 SRF cavities, providing an energy overhead of about 0.5 %. Each SRF cavity is individually powered by solid-state amplifiers of respective RF power: 7 kW (HWR), 7 kW (SSR1), 20 kW (SSR2), 40 kW (LB650) and 70 kW (HB650). The overall RF power available is compatible with a continuous wave (CW) beam operation with 2 mA current. Note that even with the 1.1% beam duty cycle for the LBNF application, the SRF linac will be operated in RF CW mode because the current cavity tuning systems are not capable of compensating the Lorentz-force detuning effect, which is up to 20 times larger, for SSR2 and LB650 cavities, than the cavity RF half-bandwidth.

The details of the SRF technology used for the SRF cavity fabrication, processing and expected performance are described in Reference [3].

The linac tunnel (Fig. 4) can accommodate two additional HB650 cryomodules for a future 1 GeV energy upgrade and an equivalent space for high intensity upgrades when a beam switchyard could be needed. This space is also used to include an emittance measurement and coupling correction section with skew quadrupoles. The end of the tunnel, beyond the switch to the BTL, includes a straight-ahead absorber, rated for 1.4 kW beam power, which will be used for low-power beam commissioning.

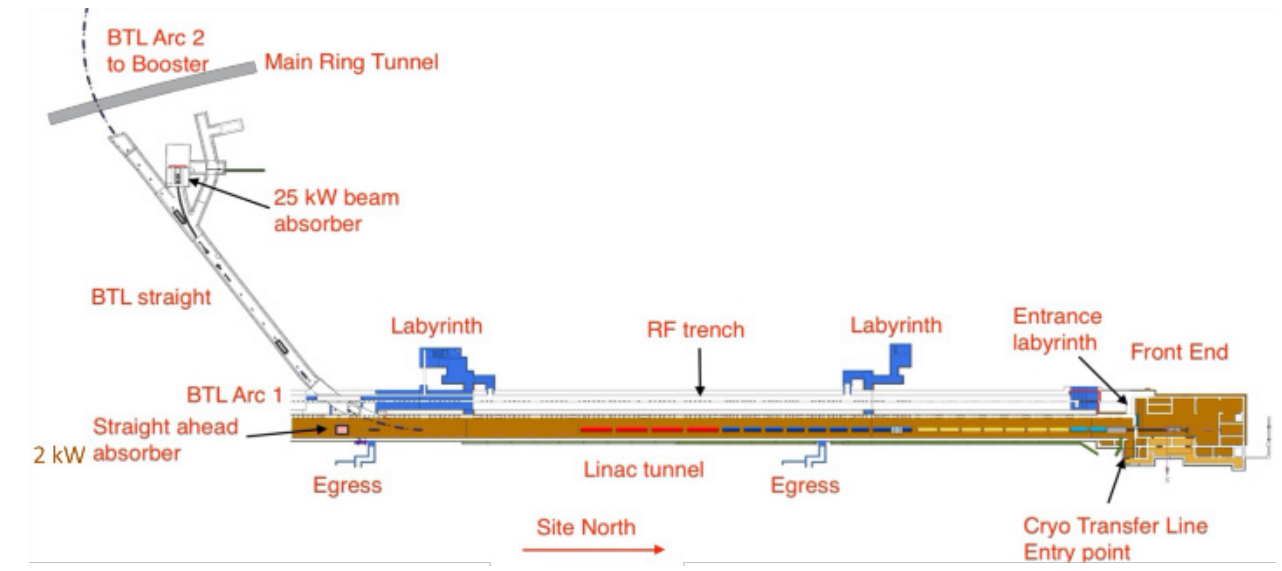

Figure 4: Layout of Linac2 and BTL.

## *The Beam Transfer Line*

The BTL includes two arcs separated by a straight section with a switch to a beam absorber line (BAL) and followed by the new Booster injection girder. The first arc (Arc-1) includes 8 dipoles forming one achromat while the second arc (Arc-2) includes 3 identical achromats. The switching of the linac beam to the BTL is directed by energizing the first dipole of Arc-1 equipped with a customized beam chamber.

The BTL straight section will include several devices to minimize the beam losses in the Booster injection area:

- a 2-stage transverse collimation system including a pair of spoiler and absorber jaws to remove about 1% beam tails on the one side of the beam corresponding to the stripping foil horizontal and vertical edges.
- A matching section composed of 6 quadrupoles to tune the transverse phase advances between the

collimator jaws and the stripping foil as well as Arc-2 optics to multiples of $2\pi$.

- A fast pulse-to-pulse switch to the beam absorber line (Fig. 4) equipped with a 25 kW-rated beam stop where the beam power is distributed over the graphite core by two horizontal and vertical sweeping dipoles. The BAL will be used to commission and tune-up the beam at full power.

The BTL ends in the Booster tunnel by transferring the $H^-$ beam to a new injection region where the $H^-$ and circulating proton beam trajectories merge through a 4-dipole symmetric chicane (ORBUMP) that creates a local vertical bump to the Booster orbit during the 550 µs time of the linac pulse injection [1].

### *Beam Instrumentation*

Linac2 is equipped with instrumentation systems, which provide beam current, position, phase, loss, length and profile measurements for Linac 2 and BTL. They must be able to operate over a wide bandwidth from few µs-long pulses for machine commissioning, to long 550 µs operational pulses, favouring 2 mA pulse current.

Two H/V pairs of Allison scanners will measure the beam emittance in the WFE. To measure beam profiles while preserving the particle-free vacuum, thirteen Laser profile monitors (LPM) will be used in the warm insertions between cryomodules.

The energy measurements, which underlie the SRF cavity phasing, will be obtained by estimating time-of-flights between successive BPMs from BPM phases measurements. For this purpose, a dedicated beam instrumentation reference line will hold 0.1° phase stability and 0.1° phase jitter at each tap at 162.5 MHZ. It is based on temperature stabilization of the phase reference line coupled with an innovative two-tone link system that uses two tones (162.5-167.5 MHz) to measure phase drifts in the phase reference cable and actively compensates for them [4].

### *Machine Protection System*

The Machine Protection System (MPS) [5] inhibits beam in cases of excessive beam loss, equipment failure, operator error or request. The goal is to limit to 20 µs the duration of a full beam loss at any location. For this, the MPS reaction time will be faster than 10 µs from detection to beam inhibit. MPS uses four beam inhibit devices all located in the 30 keV stage, namely:

- the LEBT chopper high voltage and the ion source modulator which are fast and activated at any beam interruption request,
- the ion source high voltage and the LEBT dipole which are slower and activated at interruption that requires operator intervention.

Its tight monitoring of beam current and losses will be based on ring pickups, scrapers and other intercepting devices in the WFE and an array of eight AC current transformers (ACCT) along the facility with ±1% intensity accuracy and 1.5 µA resolution.

### *Radiation Shielding*

Radiation shielding in the linac has been dimensioned for a CW beam of 1 GeV and 2 mA (2 MW beam power) assuming a beam loss of 1 W/m along the linac during normal operation, and 3 s maximum duration for accidental beam loss. The two critical devices that safely prevent the downstream propagation of beam are the ion source high voltage and the LEBT dipole power supplies.

In the BTL, radiation shielding against accidental beam loss has been calculated for a maximum beam power of 25 kW to match to the Booster injection parameters. The critical devices are the two power supplies that energize the 8 arc-1 dipoles in an alternated pattern.

Aside from potential beam halo, the main beam loss mechanism is residual gas stripping, which generates power losses well below 0.1 W/m for a CW beam of 2 mA assuming residual pressure smaller than the $7\times10^{-8}$ mbar.

## STATUS OF CONSTRUCTION

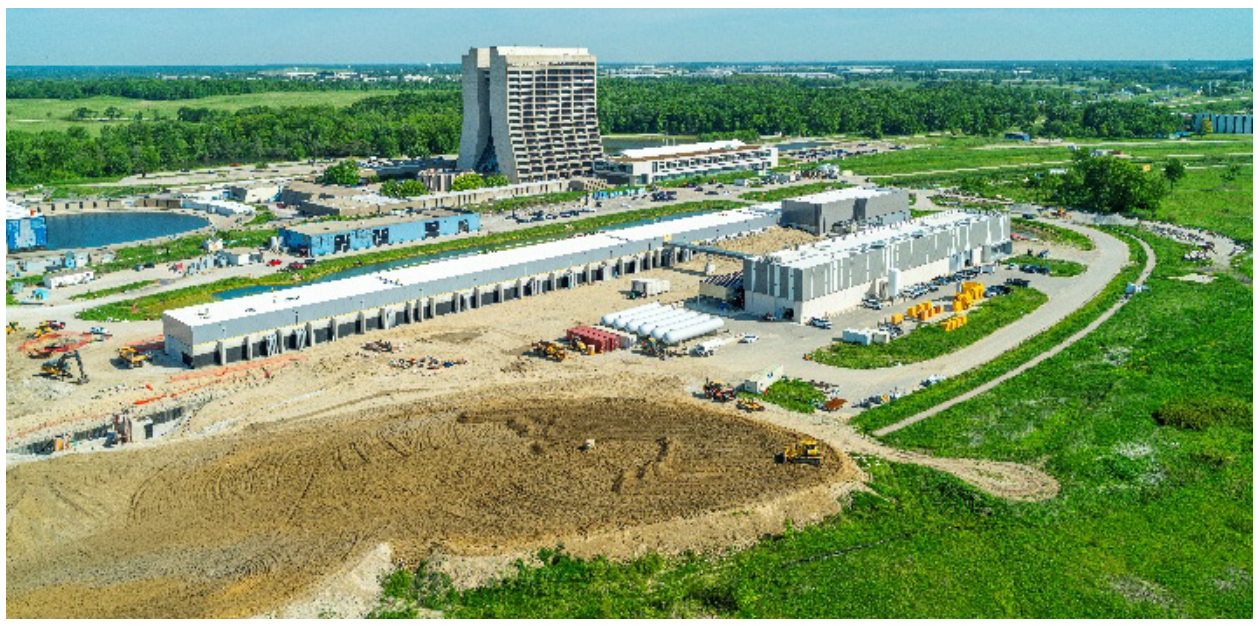

Figure 5: Overall view of the PIP-II construction site.

### *Conventional Facility*

The construction of the conventional facilities is quite advanced with authorization of usage already delivered for the cryoplant building, the high-bay building (HBB) housing the WFE beamline and services, the linac tunnel and gallery (Fig. 5) and the first part of the BTL tunnel, down to the crossing with the pre-existing Tevatron tunnel. The construction second part of the BTL tunnel connecting to the Booster tunnel is scheduled to start in the fall 2026.

Installation of cables, RF distribution, fluids and laser transfer lines is in full deployment in the HBB, the tunnel and the gallery. The first accelerator component, the RFQ, has been craned down and positioned in the HBB (Fig. 6).

### *2 kW Cryoplant*

The PIP-II cryoplant consists of one building with a compressor room and cold box room (Fig. 7), a distribution valve box (DVB) and a cryogenic distribution system (CDS). The cold box with cooling capacity of 2.5 kW at 2 K, is an Indian in-kind contribution from the Department of Atomic Energy (DAE). It is supplemented by a liquid Helium (LHe) dewar of 10,000 litres and a Helium purification and recovery system. The cryoplant has received operational readiness clearance and completion of commissioning is planned in spring 2027.

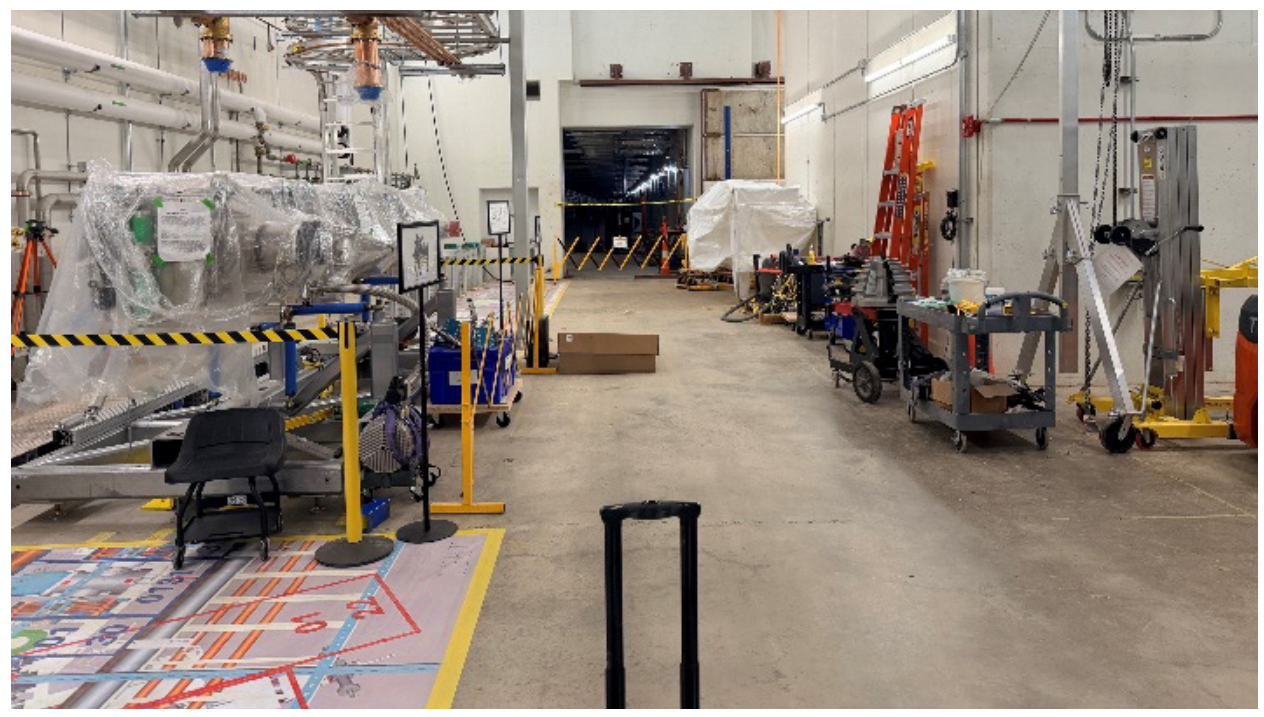

Figure 6: Injector high-bay building (HBB) with the RFQ stand and the shield door to the Linac tunnel.

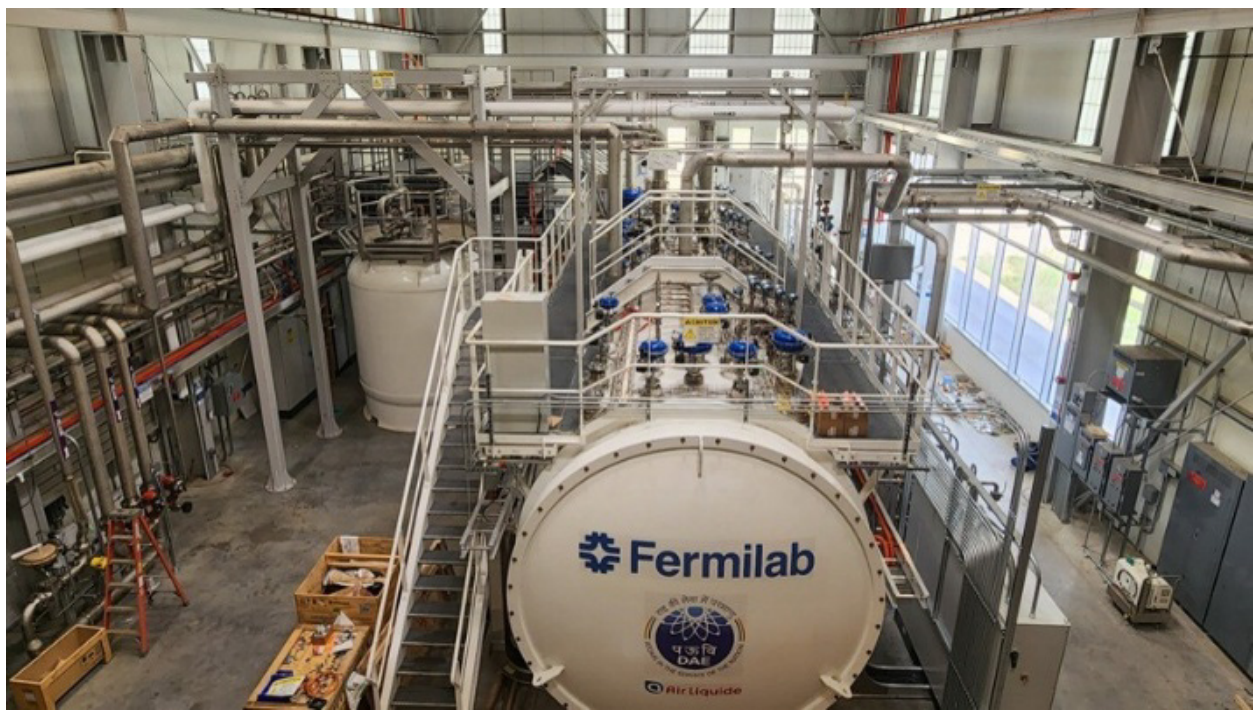

Figure 7: Cold box and LHe dewar in cryoplant building.

## SRF Cryomodules

The HWR cryomodule, which captures the beam from the RFQ at 2.1 MeV has been built by the Argonne National Laboratory (ANL) and has been tested with beam at Fermilab in 2021 [6]. The performance of the 8 cavities was validated to the specifications, but the beam did not gain energy from the first 3 cavities with the first two being 260 kHz out of tune and the coupler bias circuit of the third one not operational. The cryomodule cold mass is currently under repair at Fermilab with its beam line volume under vacuum. The tuning of the first 2 cavities was done successfully with a customized and temporary mechanical tuner (Fig. 8). When the coupler electric bias is fixed and temperature sensors added to the 8 solenoid beam tubes to improve the beam loss detection, the cryomodule will be reassembled, closed, installed in the tunnel where it will be cold tested.

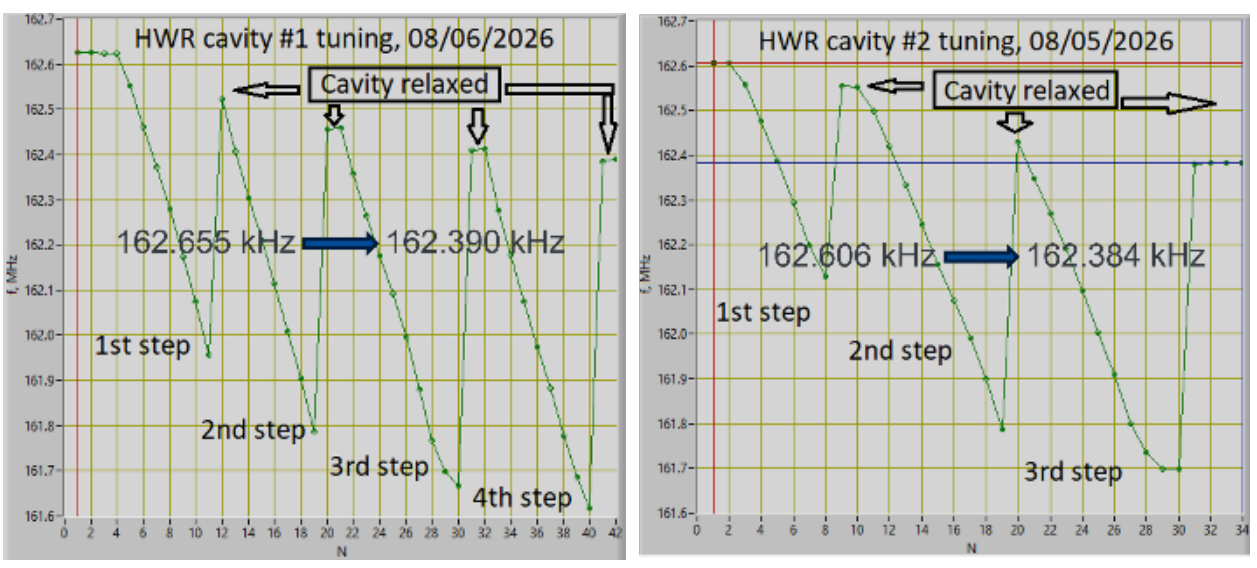

Figure 8: RF frequencies of the first two HWR cavities during their mechanical tuning operations.

The design of the four other PIP-II cryomodules is based on the common Fermilab strongback technology [3] where the string of cavities and solenoids is sitting on a strongback system kept at room temperature by its thermal anchoring to the vacuum vessel. This common design allowed to define and to follow an integrated technology roadmap for prototype and pre-production cryomodules (Fig. 9) before launching the production.

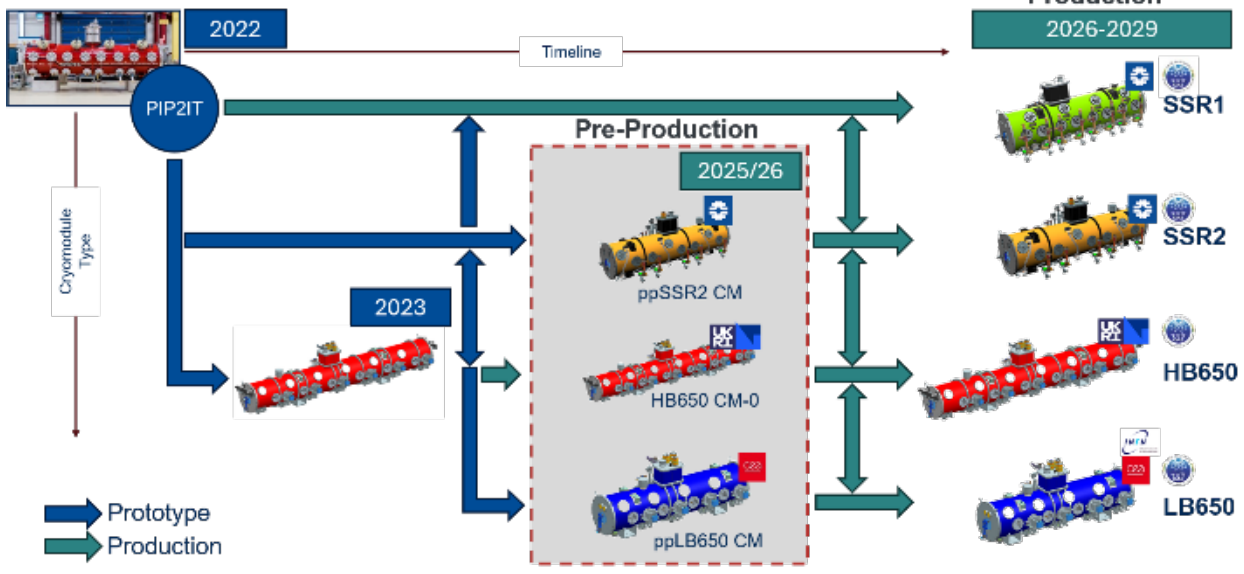

Figure 9: Cryomodule technology roadmap.

The SSR1 cryomodule prototype has been successfully tested with beam together with the HWR cryomodule [6]. It is ready for tunnel installation. Two more SSR1 cryomodules will be fabricated and assembled in Fermilab.

The pre-production SSR2 cryomodule has just been assembled at Fermilab [7] and is ready for RF testing. A major technology breakthrough, involving the Irène Joliot Curie Laboratory (Orsay, France), of the past two years has been the preparation of field-emission free spoke cavities thanks to robotic high-power rinsing system of Michigan State University. Seven more cryomodules will be fabricated and assembled by Fermilab.

The LB650 pre-production cryomodule is under assembly at CEA (France) where the cavity string has just been rolled out from the clean room. It includes four field-emission free LB650 cavities from an earlier Fermilab prototype production and preparation. Another breakthrough, shared by Fermilab [7], was the successful robotic assembly of coupler vacuum part at CEA demonstrated by individual cavity RF tests at Fermilab. Nine more cryomodules will be fabricated by INFN, CEA and Fermilab, and assembled at CEA.

The HB650 pre-production will be assembled in 2026 at UKRI (UK). More than 6 cavities, prepared and tested at UKRI, qualified without field emission. Four more cryomodules will be fabricated by DAE, UKRI and Fermilab, and assembled at UKRI and Fermilab.

The industrial production of spoke and elliptical cavities is ongoing and the first articles qualified in cold RF test, sometimes equipped with the vacuum part of the high-power coupler, with no field emission.

# CHALLENGES

Linac2 will be the first SRF injector of an RCS chain. Booster injection requires small emittances (Table 2) to achieve the expected $H^-$ beam size on the foil and the painting process within the Booster longitudinal and transverse acceptance. Linac2 lattice was optimized to decouple optics and preserve emittance (Fig. 10) in the presence of space charge and RF focusing [8] using a new in-house code HELIX [9]. Transverse coupling correction is critical for the image of the H-V collimator jaws to be mapped on the foil rectangular edges. While the transverse emittances are essentially preserved through the BTL, an RF buncher system has been designed [10] to be installed in the BTL straight section, to re-tune the longitudinal emittance and compensate the twice too large momentum spread generated by the longitudinal space charge along the BTL.

Table 2: Beam Parameters Required at Booster Injection

| Parameter | Value |
|---|---|
| Normalized transverse rms emittance, H/V | 0.22 / 0.22 mm.mrad |
| Normalized longitudinal rms emittance | 0.42 mm.mrad |
| RMS momentum spread | $2.1\times10^{-4}$ |

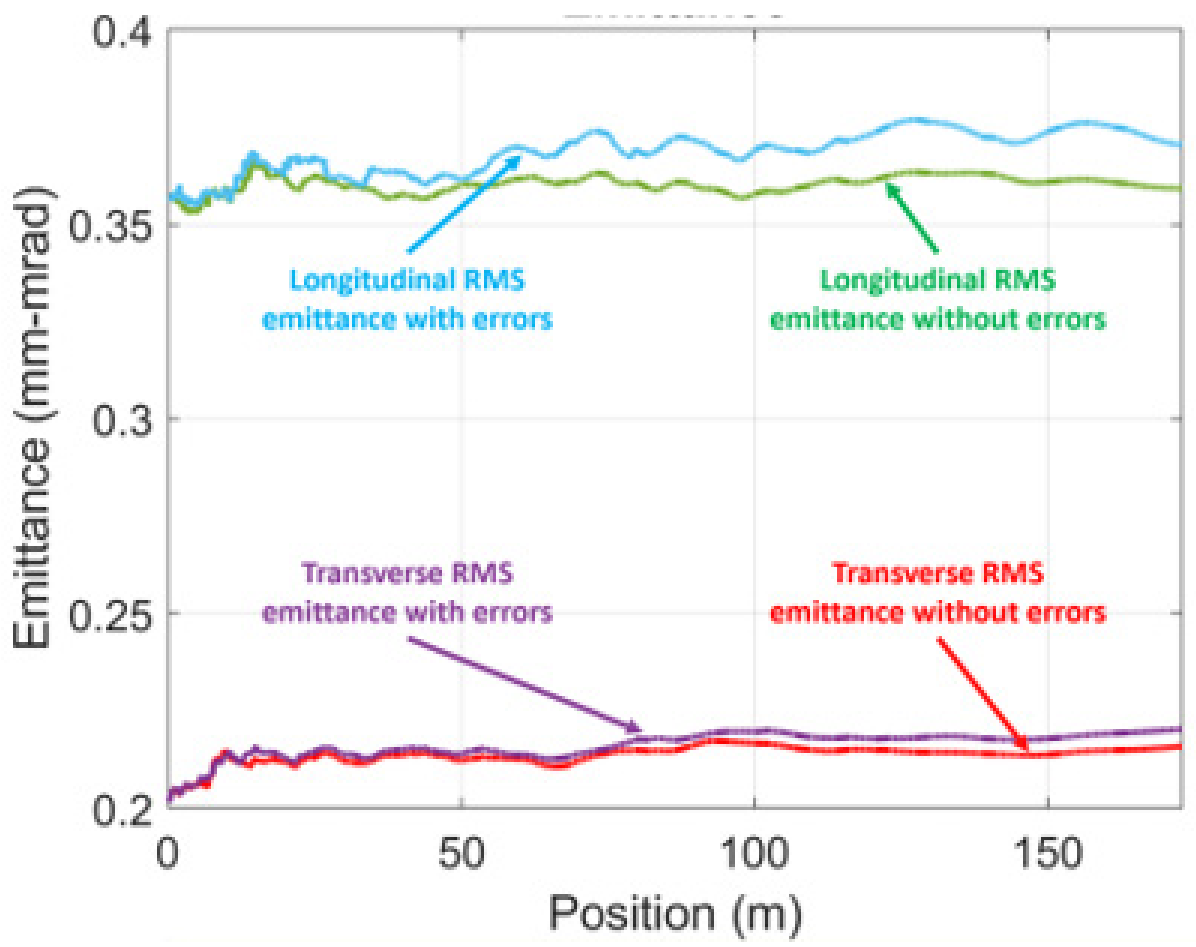

Figure 10: Calculated beam emittances along Linac2.

The Booster required $6.5\times10^{12}$ intensity per pulse will be injected in a relatively long 550 µs RF-pulse covering about 300 turns. On the Booster side, this will require flattening the orbit period during injection time by energizing dipole correctors to compensate for the 20 Hz resonant variation of the regular dipoles. On the Linac2 side, the MEBT chopper system, synchronized with the Booster RF system, will remove bunches that do not fit into the $\pm0.55\pi$ RF-phase acceptance, as well as all bunches belonging to the 3 empty Booster buckets needed for extraction, with a bunch extinction ratio smaller than $4\times10^{-4}$. The beam current at the MEBT input will thus be 3.8 mA. Together with the 0.07 % off-momentum injection, this process is designed to minimize the space-charge tune spread in the Booster by uniformizing the longitudinal charge distribution inside the remaining 81 Booster buckets.

In preparation of the beam commissioning, beam studies are conducted in the existing 400 MeV linac allowing to test and validate optics and beam tuning algorithm. These studies allow to develop and test in real conditions the high-level applications that will be used for PIP-II [11-14].

## PIP-II FACILITY UPGRADES

Besides its critical role as the injector of the chain of proton accelerators driving LBNF to 1-2 MW power-on- target, the PIP-II facility is supporting the future of Fermilab science towards physics programs requiring multi-GeV energy and/or multi MW power proton drivers.

### *Multi-GeV Energy Upgrade*

A possible 2-GeV extension of the SRF linac is shown in Fig. 11 where cryomodules included in the upgrade are shown in light green. The nominal acceleration voltage developed by the β=0.92 elliptical resonator reaches its maximum of 19.9 MV at 2.9 GeV energy compared to 15.9 MV at 800 MeV (Fig. 3). Hence HB650 cryomodules energy gain will increase along the 2 GeV extension.

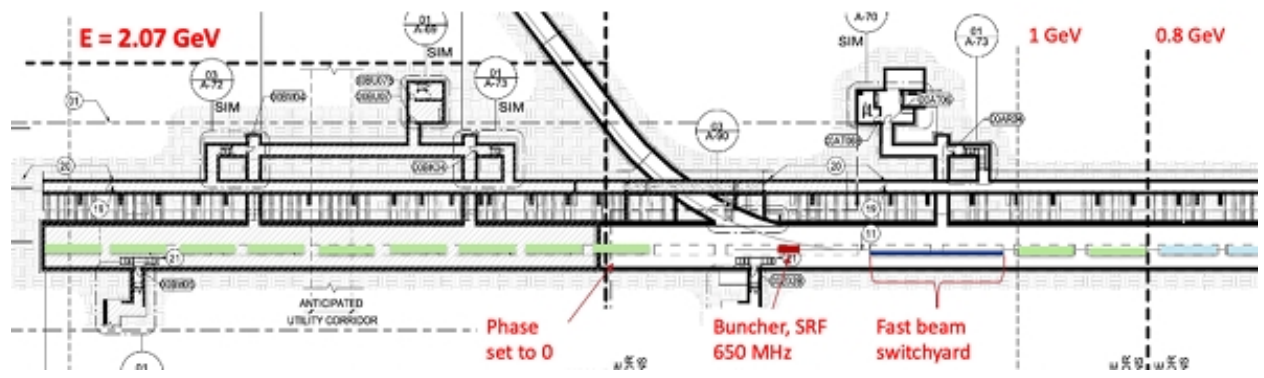

Figure 11: Layout of the 2 GeV extension.

Such a 2 GeV linac could be used as the proton driver of the Main Injector replacing the existing Booster either through a new 2-8 GeV RCS or an 8 GeV linac built from 1.3 GHz SRF cryomodules with β=1 elliptical cavity.

### *CW Upgrade*

PIP-II SRF Linac is CW compatible by design. This means that all accelerator components can operate in CW mode and that the only restriction to it is that no system is in place to dispose of a MW-range beam power. While the SRF linac will operate in RF CW mode for the 1.1% beam duty cycle of the LBNF application, some hardware of the WFE will need to be upgraded mostly for managing higher thermal loads. One example is the insufficient cooling capacity of the driver transistors of the MEBT kickers currently limited to an average switching frequency of 500 MHz. Another concern is the ability to manage higher beam losses due to longitudinal beam tails at the output of the RFQ.

CW operation would allow Linac2 to serve multiple beam lines delivering high intensity beam to different users, for example by using a 3-way switching system based on RF separators at one fourth of the 162.5 MHz bunch repetition rate. It could also serve as an MW-power facility.

## CONCLUSION

The PIP-II facility will enrich Fermilab a modern state-of-the-art SRF linac, now renamed Linac2, which will be the basis for the ambitious neutrino physics program of the LBNF/DUNE international experiment. Linac2 is expected to operate over at least 40 years and will allow Fermilab to remain on the forefront of accelerator developments and physics programs for the next decades.